\documentclass[conference]{IEEEtran}

\usepackage{amsmath,amssymb,amsfonts}
\usepackage{cite}
\usepackage{hyperref}
\usepackage[utf8]{inputenc} 
\usepackage[T1]{fontenc}
\usepackage{algorithmic}
\usepackage{tikz}
\usepackage{listings}
\usepackage{tcolorbox}
\usepackage{color}

\newcommand{\bx}{{\mathbf{x}}}

\newcommand{\bX}{{\mathbf{X}}}
\newcommand{\bY}{{\mathbf{Y}}}
\DeclareMathOperator{\supp}{\mathrm{supp}}

\definecolor{keywordcolor}{rgb}{0.7, 0.1, 0.1}   
\definecolor{tacticcolor}{rgb}{0.0, 0.1, 0.6}    
\definecolor{commentcolor}{rgb}{0.4, 0.4, 0.4}   
\definecolor{symbolcolor}{rgb}{0.0, 0.1, 0.6}    
\definecolor{sortcolor}{rgb}{0.1, 0.5, 0.1}      
\definecolor{attributecolor}{rgb}{0.7, 0.1, 0.1} 

\newtheorem{theorem}{Theorem}[section]
\newtheorem{definition}[theorem]{Definition}
\newtheorem{example}{Example}[section]
\newtheorem{remark}{Remark}[section]

\def\BibTeX{{\rm B\kern-.05em{\sc i\kern-.025em b}\kern-.08em
		T\kern-.1667em\lower.7ex\hbox{E}\kern-.125emX}}

\title{Aligning Information Theory
and Formal Verification: 
A Lean Formalization of Rate–Distortion Theory}

\author{Fubin Yan, Ruxuan Dong, Ruoxuan Shi, Duolei Wang, and  Kenneth W. Shum \\
School of Science and Engineering\\
The Chinese University of Hong Kong, Shenzhen\\
Shenzhen, China.}

\begin{document}
\maketitle
\pagestyle{plain}
\begin{abstract}
Large language models have recently demonstrated an ability to generate substantial mathematical proofs and formalizations, including machine‑checked arguments for major open problems and classical theorems. These developments highlight a growing workflow in which AI systems search for proofs while proof assistants certify their correctness. Within information theory, however, formalized libraries remain limited, and existing formalized theorems focus almost exclusively on Shannon’s foundational lossless theorems.
This paper presents the first Lean formalization of core components of rate–distortion theory, a central framework for lossy compression and modern machine‑learning analysis. We identify the alignment challenges that arise when translating information‑theoretic definitions, such as distortion measures, typicality and achievability, into a proof‑assistant setting, and we develop a reusable library of definitions and lemmas faithful to textbook conventions. Our formalization provides a foundation for future AI‑assisted proofs in information theory and enables subsequent mechanized developments to build on a coherent, verified infrastructure rather than generating long proofs from scratch.
\end{abstract}


\section{Introduction}



Recent breakthroughs in AI-assisted mathematics have demonstrated that large language models can now produce substantial, technically sophisticated proofs, either with human guidance or even from a single prompt. Two landmark examples are a solution attempt for the Navier–Stokes equation and a complete computer formalization of Fermat’s Last Theorem. In September 2026, an internal OpenAI model generated a 166‑page proof of Statements C and D from the Clay Mathematics Institute’s millennium prize announcement~\cite{OpenAINavierStokes2026}, and verified its correctness in Lean. In the same month, Anthropic’s Claude produced a full formalization of Fermat’s Last Theorem, comprising 13.4 million lines of Lean code and 29,500 intermediate lemmas~\cite{AnthropicFLT2026}.

\begin{table}
\begin{center}
\caption{Recent AI-assisted Works in Information Theory}
\label{table:related_works}
\begin{tabular}{c|c|c}
Topic & Lean formalized & Reference \\ \hline 
Cyclic Hadamard matrix & no & \cite{OAI:The-circulant-Hadamard-conjecture-September-23-2026} \\
and Barker sequence conjecture & & \\ \hline
Three conjectures about doubly  & yes & \cite{Pichler2026} \\
symmetric binary source & & \\ \hline 
Weak Hellinger inequality for & yes & \cite{Durcik2026weak} \\
noisy Boolean channels && \\ \hline 
Upper bound on binary & yes&\cite{Soysal2026new} \\
deletion channel capacity && \\ \hline 
Sub-optimality of Marton & no & \cite{Huang2026} \\
inner bound && \\ \hline 
Classification of $(n,4)$cdoe  & yes & \cite{YangDongN4Code} \\
for binary symmetric channel && \\ \hline 
Improvement of MRRW bound & yes & \cite{OpenAI10Advances} \\ \hline 
Improved lower bound on & yes & \cite{Itty2026improved, Tandon2026, Protti2026Shannon} \\
zero-error capacity of $C_{n}$ && \\ \hline 
\end{tabular}
\end{center}
\end{table}


These developments illustrate an emerging workflow for AI-assisted mathematics. Before an AI system can effectively search for a proof, the relevant definitions, hypotheses, and target statements must be expressed with complete precision. After a proof is found -- especially one that is long or LLM generated -- formal verification provides a mechanism to certify that no hidden gaps remain in the formal proof relative to the stated definitions and hypotheses. Table~\ref{table:related_works} summarizes several recent results in information theory obtained with assistance from AI systems.

For elementary combinatorial or algebraic problems, the computer formalization step is often straightforward. In contrast, many problems in information theory require a substantial layer of background definitions and customary conventions before the statement itself becomes mathematically meaningful. Concepts such as entropy, mutual information, KL divergence, typicality, and asymptotic achievability form a specialized language for reasoning about communication systems and statistical learning. Ensuring that formal definitions and theorem statements faithfully capture the intended mathematics is known as the {\em alignment problem}.
Lean verifies proofs against the formal definitions and
statements, but their correspondence with the intended
mathematics requires separate checking.
Bastounis, Circelli, and Hansen~\cite{Bastounis2026}
discuss mismatches between natural-language arguments
and their Lean formalizations.

In our development, we checked the formal definitions
and theorem statements against the intended mathematics
before using AI assistance to complete the proofs.
This method allows us to benefit from the speed of
AI-assisted code generation while reducing the risk
of mismatches with the intended mathematics.

Despite recent advances in auto-formalization, the Lean proofs produced in the works listed in Table~\ref{table:related_works} are not easily reusable in other projects. For example, the proof of the improved MRRW bound in~\cite{OpenAI10Advances} is so long that reading it line by line is infeasible for humans. If a new AI-generated proof is produced and one wishes to verify it in Lean, current AI models typically start from scratch and generate yet another lengthy proof. A well‑designed library of definitions and theorems in Lean, that is aligned with textbook conventions and standard information-theoretic practice, is therefore essential for simplifying future formalization efforts and enabling reuse across projects.

Currently, the mathematical library of Lean contains some basic results in information theory, such as KL divergence and Hamming distance. However, the library is not complete and the development is on-going. In a companion of the proof of polynomial Freiman--Ruzsa conjecture~\cite{gowers2025conjecture}, Tao and his collaborators
need the Ruzsa distance in their project, and for this purpose, they implemented Shannon entropy and mutual information for up to three random variables~\cite{PFRLean}.
Luccioli in his master's thesis formalized $f$-divergences, and the associated data processing inequality in Lean~\cite{Luccioli_thesis}. 
Concentration inequalities, Rademacher complexity and Dudley’s entropy integral theorem for sub-Gaussian processes are implemented in~\cite{sonoda2026Lean, zhang2026empirical}, providing a formal foundation for statistical learning theory.
A more comprehensive project on building a Lean library for machine learning research was initiated by Degenne and Rauber~\cite{Degenne2026Lean}, with the aim of  analyzing machine learning algorithms in Lean.


Rate--distortion theory is a central branch of classical information theory, characterizing the minimum communication rate required to represent a source subject to a prescribed distortion constraint. In recent years, it has also become an important tool in the analysis and design of neural networks and machine‑learning systems. Since Shannon’s foundational work and its systematic development in standard texts such as Berger~\cite{Berger71}, Viterbi and Omura~\cite{Viterbi2009}, Cover and Thomas~\cite{CoverThomas}, McEliece~\cite{McEliece2022}, and Yeung~\cite{Yeung}, rate--distortion arguments have become indispensable in lossy compression, quantization, statistics, learning theory, and network information theory.

Although rate--distortion theory plays a central role in information theory, it has not yet been formalized in mainstream proof assistants. Existing computer formalization in information theory have focused primarily on Shannon’s foundational lossless results, and have been carried out mainly in Isabelle/HOL or the Rocq proof assistant. Mhamdi, Hasan, and Tahar formalized entropy measures in HOL~\cite{Mhamdi2011}. In parallel, Affeldt, Hagiwara, and Senizergues developed a substantial formalization of core Shannon theory—including coding theorems and typical‑sequence arguments—first reported in~\cite{Affeldt2012} and later expanded in~\cite{Affeldt2014}. This line of work was consolidated into the \texttt{infotheo} library, which provides a Rocq infrastructure for classical coding theorems~\cite{CoqInfotheo}, and has supported further case studies such as the formalization of the binary symmetric erasure channel~\cite{Nakano2016}. In another direction, a Lean formalization of quantum information theory has recently been developed in~\cite{kasaura2026lean}.

This paper fill this gap by initiating a Lean formalization of the core of rate--distortion theory for finite alphabets. We introduce the information measures, including entropy, KL divergence, and mutual information, using Mathlib's existing infrastructure. We then establish the supporting inequalities and probabilistic estimates, and use them to prove the direct and converse results of the rate--distortion theorem. The direct result establishes achievability for $R>R_I(D)$, with every code satisfying the prescribed rate bound, while the converse holds for all distortion thresholds. Under an asymptotic rate constraint, we also prove achievability at $R=R_I(D)$ whenever $R_I(D)$ is finite. We also establish the existence of optimal test channels and the right continuity of the information rate-distortion function. This work also provides some reusable definitions and lemmas in information theory built on Mathlib. The long-term objective is to build a library for reusable formalization that can be extended toward more advanced topics, such as multi-terminal problems, information geometry, and statistical learning theory.

The remainder of this paper is organized as follows. We review the problem formulation of rate-distortion problem for finite alphabet in Section~\ref{sec:problem_formulation}, and  define the basic information measures, such as entropy, KL divergence and mutual information 
in Section~\ref{sec:lean-information-measures}. In Section~\ref{sec:lean-rate-distortion-function}, we define the information rate-distortion function as the optimal value of an optimization problem with mutual information as the objective function. After setting up these prerequisites, we formulate the operational rate-distortion problem in Section~\ref{sec:lean-formulation} and state the direct and converse parts of the rate-distortion theorem in Lean. Section~\ref{sec:lean-proof} presents the formal proofs and discusses the reusable components developed along the way. Finally, Section~\ref{sec:conclusion} concludes the paper.

\section{Problem formulation}
\label{sec:problem_formulation}

We first recall the problem statement of the rate--distortion problem in this paper. We will see later how the mathematics are translated into Lean.

We consider lossy compression with a fixed block length~$n$ and finite alphabet.  The source data $\bX^n$ is encoded to an element in $\{1,2,\ldots, M\}$, which is sent to the decoder with no error. The decoder then computes the estimate $\hat{\bX}^n$. We assume that random variables $X_k$ and $\hat{X}_k$, for $k=1,2,\ldots, n$, take value in $\mathcal{X}$ and~$\hat{\mathcal{X}}$. In general, $\mathcal{X}$ may not be the different from $\hat{\mathcal{X}}$

\begin{figure}
	\begin{center}
		\begin{tikzpicture}[scale = 0.9]
			\draw
			(0,0) node[minimum height = 1cm, rectangle, draw] (source) {Source}
			(3,0) node[minimum height = 1cm, rectangle, draw] (C) {Encoder}   
			(7,0) node[minimum height =1cm, rectangle, draw](D){Decoder} 
			(9,0) node (sink) {};
			\draw[->](source) -- (C) node[midway, above] {$\bX^n$} ;
			\draw[->](C) -- (D) node[midway, above] {$enc(\bX^n)$} ;
			\draw[->](D) -- (sink) node[midway, above] {$\bY^n$} ;
		\end{tikzpicture}
		\caption{Rate distortion code.}
		\label{fig:rate_distortion_model}
	\end{center}
\end{figure}
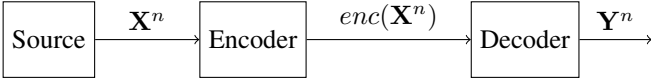

\begin{definition}
An \emph{$(n,M)$ rate distortion code} consists of an encoder 
$$enc: \mathcal{X}^n \rightarrow \{1,2,\ldots, M \},$$
and a decoder 
$$dec: \{1,2,\ldots, M\}\rightarrow \hat{\mathcal{X}}^n.
$$ 
A sequence in the image of the decoder function $dec$  is called a \emph{codeword}. The \emph{rate} of an $(n,M)$ code is defined as $R= (\log_2 M)/n$. (see Fig.~\ref{fig:rate_distortion_model})
\end{definition}

We aim at minimizing a per-letter {\em distortion function}
$$
d: \mathcal{X} \times \hat{\mathcal{X}} \rightarrow \mathbb{R}_{\geq 0}.
$$
A distortion function $d$ is said to be {\em bounded} if 
$$\max_{x,\hat{x}} d(x,\hat{x}) < \infty.
$$
We extend the  distortion function to vector input: 
\begin{equation}
d(\bx^n,\hat{\bx}^n) \triangleq  \sum_{k=1}^n d(x_k,x_k')
\label{eq:distortion_vector}
\end{equation}
for $\bx^n=(x_1,\ldots, x_n) \in \mathcal{X}^n$ and $\hat{\bx}^n = (\hat{x}_1,\ldots, \hat{x}_n)$  in $\hat{\mathcal{X}}^n$.

\begin{example} (Mean-squared error) The distortion measure
	$$
	d(x, \hat{x}) = (x-\hat{x})^2
	$$
defined for $\mathcal{X}=\hat{\mathcal{X}} \subseteq \mathbb{R}$ is called the {\em mean-squared distortion}. The corresponding distortion between two real vectors $\bx^n$ and $\hat{\bx}^n$ is the squared Euclidean distance between $\bx^n$ and $\hat{\bx}^n$. 

The $k$-means clustering is an example of rate distortion code with mean-squared error. The source distribution $p(\bx^n)$ is the empirical distribution of the data set. 
\end{example}

When $\mathcal{X}$ and $\hat{\mathcal{X}}$ are finite, we can represent the distortion function by an $|\mathcal{X}|\times |\hat{\mathcal{X}}|$ matrix
$$
\begin{bmatrix}
d(x_i,\hat{x}_j) 
\end{bmatrix}_{i=1,\ldots, |\mathcal{X}| ,\  j=1,2,\ldots, |\hat{\mathcal{X}}|}
$$
This matrix is called the {\em distortion matrix}.

\begin{example}
Suppose the alphabet $\mathcal{X}$ is $\{0,2\}$, and the output of decoder can take values in $\hat{\mathcal{X}} = \{0,1,2\}$. We define a distortion function by matrix
$$
\begin{bmatrix}
	0 & 1 & 2 \\
	2 & 1 & 0
\end{bmatrix}.
$$
This distortion between $x$ and $\hat{x}$ is the absolute value $|x-\hat{x}|$.
\end{example}

In the general formulation of rate distortion, the components of input sequence $\bX$ may not be independent. We can assume that $\bX$ follows a general joint distribution $p(\bx^n)$ on~$\mathcal{X}^n$. The average distortion of an $(n,M)$ code is 
\begin{align*}
 E[d(\bX^n, Q(\bX^n))] 
= \sum_{\bx^n \in \mathcal{X}^n} p(\bx^n) \cdot  d(\bx^n, Q(\bx^n)).
\end{align*}

\begin{definition}
A rate distortion-pair $(R,D)$ is called \emph{achievable} if there exists a sequence of $(n,M_n)$ rate distortion codes with increasing length, such that the rates are all less than or equal to $R$, and 
\begin{equation}
\limsup_{n\rightarrow \infty} E[\frac{1}{n} d(\bX^n, \hat{\bX}^n)] \leq D.
\label{eq:rate_distortion_average}
\end{equation}
The random vector $Y^{n+1}$ is the estimate of $X^{n+1}$. The {\em rate-distortion region} is the closure of the set of all achievable rate-distortion pairs.
 \label{def:rate_distortion_achievable}
\end{definition}

Cover and Thomas in their textbook \cite{CoverThomas}, define achievability by
$$
\lim_{n\rightarrow \infty}  E[\frac{1}{n} d(\bX^n, \hat{\bX}^n)] \leq  D.
$$
This definition is simpler, but requires that the limit exists. In contraast, the limsup of a sequence of numbers always exists, and hence \eqref{eq:rate_distortion_average} is more flexible. Intuitively speaking, achievability means that we can get arbitrarily close to the point $(R,D)$ by designing rate distortion codes of long block lengths. 


\begin{remark}
In the literature, there is an alternate criterion for achievability, obtained by replacing  \eqref{eq:rate_distortion_average} by
$$
\Pr (\frac{1}{n} d(\bX^n, \hat{\bX}^n) \leq  D + \epsilon ) \geq 1 - \delta,
$$
for $\epsilon>0$ and $\delta>0$. That is, if we generate input sequence $\bX^n$ randomly according to distribution $p(\bx^n)$, then the distortion measure is less than $D+\epsilon$ with probability at least $1-\delta$.
Lossy compression with this objective function is sometime called ``source coding with a fidelity criterion,'' and it is treated in~\cite{Yeung}. 
\end{remark}

\subsection{Information rate-distortion function}

We consider the simplest probability model and suppose that
the source symbols $X_1,\ldots, X_n$ are generated independently from a known distribution $p_X(x)$. We use the notation $X$ to denote a generic random variable that has the same distribution as $X_1,\ldots, X_n$.

Every conditional distribution $p(\hat{x} | x)$ induces a joint distribution of $X$ and $\hat{X}$, 
$$
p(x,\hat{x}) = p(x) p(\hat{x}| x).
$$
Using the joint distribution we can compute the expected distortion $E[d(X,\hat{X})]$ and mutual information $I(X;\hat{X})$.

\begin{definition}[Information rate distortion function] For $D\geq 0$, define
		\begin{equation}
			R_I(D) \triangleq \min_{\substack{p(\hat{x}|x) \\ E[d(X,\hat{X})\leq D ]}} I(X;\hat{X}).
		\label{eq:rate_distortion_function}
		\end{equation}
The minimum is taken over all conditional distribution function $p(\bx|x)$ such that $E[d(X,\hat{X})]\leq D$. \label{def:rate_distortion_function}
\end{definition}

Alternately, we may write \eqref{eq:rate_distortion_function} as
$$
R_I(D) \triangleq \min_{\substack{\hat{X} \\ E[d(X,\hat{X})]\leq D}} I(X;\hat{X}),
$$
with the minimum over all random variables $\hat{X}$ that are jointly distributed with~$X$, such that the expected distortion is less than or equal to~$D$. Since $p(x)$ is fixed, we need to consider all conditional distribution function $p(\hat{x}|x)$ that satisfy the average distortion constraint. The conditional distribution function $p(\hat{x}|x)$ is sometime called the {\em test channel}. 

When $\mathcal{X}$ and $\hat{\mathcal{X}}$ are finite, the test channel is specified by $|\mathcal{X}| \cdot |\hat{\mathcal{X}}|$ variables. Since $I(X;\hat{X})$ is convex as a function of $p(\hat{x}|x)$, and $E[d(X,\hat{X})]\leq D$ is a convex constraint, the minimum in \eqref{eq:rate_distortion_function} is well-defined, and is achieved by at least one choice of $p(\hat{x}|x)$. 

\begin{definition}
Let
$$
D_{\max} \triangleq \min_{\hat{x}\in \hat{\mathcal{X}}} E[d(X,\hat{x})]
$$
denote the least average distortion over all $\hat{x}\in \hat{\mathcal{X}}$, and 
$$
D_{\min} \triangleq \sum_{x\in\mathcal{X}} p(x) \min_{\hat{x}} d(x,\hat{x})
$$
be the smallest attainable average distortion.
\end{definition}

We can achieve distortion $D_{\max}$ with zero rate. The decoder always outputs the symbol $\hat{x}^*$ that minimizes the average distortion $E[d(X,\hat{x})]$. By letting $\hat{X}$ to be a constant random variable that equals $\hat{x}^*$ with probability 1 
in \eqref{eq:rate_distortion_function}, we see that $R_I(D)=0$ for all $D\geq D_{max}$
(because $I(X;\hat{X}) \leq H(\hat{X})=0$).

On the other extreme, suppose the encoder sends all the symbols $X_1, X_2,\ldots$ to the decoder. In this scenario,  the decoder can minimize the distortion by returning the symbol
$$
g(x) = \arg \min_{\hat{x}} d(x,\hat{x}).
$$
The output of the decoder is a deterministic function of $X$. The mutual information $I(X;\hat{X})$ equals
$$
I(X;\hat{X}) = H(\hat{X}) - H(\hat{X}| X) = H(\hat{X}).
$$
Since $X$ is distributed according to $p(x)$, we can achieve average distortion $D_{\min}$ with rate $H(\hat{X})$. Furthermore, since $H(\hat{X})\leq H(X)$, we get
$$R_I(D_{\min}) \leq H(X).
$$
Since this is the smallest possible distortion, we have $R(D)=\infty$ if $D < D_{\min}$. 

\begin{example}
Consider a binary source with distribution
$$
p(0)=0.3 \text{ and } p(1) = 0.7,
$$
and  distortion matrix 
$$
\begin{bmatrix}
0 & 2 \\
3 & 0 
\end{bmatrix}.
$$
If the decoder returns the constant 1, the average distortion is $0.3(2)= 0.6$. If the decoder returns the constant 0, the average distortion is $0.7(3)=2.1$. Hence, the least average distortion is $D_{\max}= 0.6$.
\end{example}

If we increase distortion $D$, the feasible set corresponding to $E[d(X,\hat{X})]\leq D$ becomes larger, and the minimum value either becomes smaller or does not change. Therefore, the information distortion function is non-increasing. In fact, one can prove that $R_I(D)$ is strictly decreasing for $D_{\min}<D< D_{\max}$.

\begin{theorem} Suppose alphabet $\mathcal{X}$ is finite. For $D_{\min} < D < D_{\max}$, the information rate distortion function $R_I(D)$ is strictly decreasing.
\end{theorem}

\begin{theorem}
The function $R_I(D)$ is convex for $D\geq D_{\min}$.
\end{theorem}

The rate distortion theorem states that the information distortion function is the solution to the rate distortion problem, providing the fundamental tradeoff between rate and distortion. The rate-distortion region is precisely the points above the curve $R=R_I(D)$.

	\begin{theorem} 
\label{thm:rate_distortion_theorem}
Suppose alphabet $\mathcal{X}$ and $\hat{\mathcal{X}}$ are finite and the distortion function $d(x,\hat{x})$ is upper bounded by a constant~$d_0$.
For any  probability distribution $p(x)$ defined on $\mathcal{X}$, we have
		\begin{enumerate}
			\item If $R > R_I(D)$, the rate distortion pair $(R,D)$ is achievable;
			
			\item If $R < R_I(D)$, the rate distortion pair $(R,D)$ is not achievable. 
		\end{enumerate}
	\end{theorem}

\section{Lean Formalization of Information Measures}
\label{sec:lean-information-measures}

We work with finite source and reproduction alphabets \texttt{X} and \texttt{Y}, which are expressed in Lean by the type class \texttt{Fintype}. For simplicity, we will treat finite types as finite sets. We impose that equality is decidable by \texttt{DecidableEq}. 

\begin{lstlisting}[language=lean, frame=single]
-- finite alphabets
variable {X Y : Type*} [Fintype X] [Fintype Y] [DecidableEq X] [DecidableEq Y]
\end{lstlisting}

Lean’s Mathlib represents a probability mass function on a finite type \texttt{X} as \texttt{PMF X}. A value \texttt{P : PMF X} is a function assigning to each \texttt{x} of type \texttt{X} a nonnegative real number summing to $1$. Internally, probabilities are stored as \texttt{ENNReal} (the extended nonnegative reals $[0,\infty]$). When calculating entropy, we typically work in the real number system because we need logarithms; thus we often convert probabilities to reals using \texttt{toReal}.

\subsection{Entropy}

The Shannon entropy is defined in the standard way:
\[
H(P) = -\sum_{x\in \texttt{X}} P(x)\log P(x).
\]
Here and throughout the paper, $\log$ denotes the natural logarithm.
In Lean, the sum over a finite type is written using the summation notation \texttt{$\sum$ x, ...}. The implementation below follows the formula directly. Here, \texttt{Real.negMulLog p} denotes $-p\log p$.

A technical point is the convention at $P(x)=0$. Mathlib defines \texttt{Real.log 0 = 0}, which makes the term $-0\cdot \log 0$ evaluate to $0$, matching the standard information-theoretic convention.

\begin{lstlisting}[language=lean, frame=single]
-- Entropy function
noncomputable def shannonEntropy (P : PMF X) : ℝ :=
  ∑ x : X, Real.negMulLog (P x).toReal
  
-- We use the convention that log 0 = 0 in entropy calculations
example : Real.log 0 = 0 := by
	exact Real.log_zero
	
-- convention that 0 * log 0 = 0
example : - (0 : ℝ) * Real.log 0 = 0 := by
	simp [Real.log_zero]
\end{lstlisting}

{\em Remark:} Some definitions in Mathlib (in particular those involving real logarithms, infima, and limit operations) are marked \texttt{noncomputable}: they are mathematically well-defined but not intended to be executed as programs. This is standard for developments that focus on theorem proving rather than computation. The code excerpts below are taken from a \texttt{noncomputable section}.

\subsection{Joint entropy}

A joint distribution on $X\times Y$ is represented as \texttt{PMF (X × Y)}. The joint entropy is just entropy on the product alphabet; 
it can be expressed by the following wrapper.

\begin{lstlisting}[language=lean, frame=single]
-- Joint entropy function
noncomputable def jointEntropy (P_XY : PMF (X × Y)) : ℝ :=
	shannonEntropy P_XY
\end{lstlisting}

\subsection{KL divergence and mutual information}

Given distributions $P$ and $Q$ on the same alphabet, the Kullback--Leibler divergence is
\[
D(P\|Q) = \sum_{x\in\supp(P)} P(x)\log\frac{P(x)}{Q(x)},
\]
where $\supp(P)$ represents the support of distribution $P$.
This formula applies when $\supp(P)\subseteq\supp(Q)$;
otherwise, the divergence is $+\infty$.
In formalization, we need to handle both the case $P(x)=0$, where the summand is taken to be $0$, and the case $P(x)>0$ but $Q(x)=0$, where the divergence is $+\infty$.
The following code uses Mathlib's definition of KL divergence for the measures associated with $P$ and $Q$.

\begin{lstlisting}[language=lean, frame=single]
/-- Definition of KL divergence -/
noncomputable def kl_divergence (P Q : PMF X) : ENNReal :=
  letI : MeasurableSpace X := ⊤
  InformationTheory.klDiv P.toMeasure Q.toMeasure
\end{lstlisting}

To define mutual information from a joint distribution $P_{XY}$, we use the identity
\[
I(X;Y) = D\!\left(P_{XY}\,\middle\|\,P_X\otimes P_Y\right),
\]
where $P_X$ and $P_Y$ are the marginals of $P_{XY}$ and $P_X\otimes P_Y$ is the product distribution.

In Lean, the marginals are obtained via \texttt{map} applied to \texttt{Prod.fst} and \texttt{Prod.snd}. The product distribution is built using the \texttt{PMF} monadic operations:
\texttt{bind} corresponds to sampling $x\sim P_X$ and then sampling $y\sim P_Y$ independently, and \texttt{map} packages the pair \texttt{(x,y)}.

\begin{lstlisting}[language=lean, frame=single]
/-- definition of product distribution Q_XY = P_X ⊗ P_Y -/
def product_pmf (P_X : PMF X) (P_Y : PMF Y) : PMF (X × Y) :=
  P_X.bind (fun x => P_Y.map (fun y => (x, y)))
	
/-- Mutual information based on KL divergence -/
noncomputable def mutual_information
    (P_XY : PMF (X × Y)) : ENNReal :=
  let P_X := P_XY.map Prod.fst -- marginal p x
  let P_Y := P_XY.map Prod.snd -- marginal p y
  -- Mutual information is given by D(P_XY || Q_XY).
  kl_divergence P_XY (product_pmf P_X P_Y)
\end{lstlisting}

The above Lean  program is a compact and faithful encoding of the textbook definitions: entropy is a finite sum over the alphabet, KL divergence uses Mathlib’s measure-theoretic definition, and mutual information is defined as KL divergence to the product of marginals.

\section{Defining the Rate--Distortion Function}
\label{sec:lean-rate-distortion-function}

In this section we explain how to define the \emph{rate--distortion function} using the standard ``test channel'' (conditional distribution) formulation. Fix a source distribution $P$ on \texttt{X} and a distortion function $d : \texttt{X}\times \texttt{Y}\to[0,\infty)$. A \emph{test channel} is a conditional distribution $W(\cdot\mid x)$ on \texttt{Y} for each element $x$ in~\texttt{X}. In Lean, we represent such a conditional distribution as a function \texttt{W : X → PMF Y}.

\subsection{Distortion function and single-letter expected distortion}

We model the distortion measure as an \texttt{NNReal}-valued function, so that each distortion value is finite and nonnegative. Since the alphabets are finite, the distortion measure is bounded. The expected single-letter distortion under $P$ and $W$ is
\[
\mathbb{E}[d(X,Y)] = \sum_x P(x)\sum_y W(y\mid x)\, d(x,y).
\]
This becomes a nested finite sum in Lean.

\begin{lstlisting}[language=lean, frame=single]
-- Distortion function
def Distortion := X → Y → NNReal
-- Single-letter expected distortion
noncomputable def expected_distortion_single
    (P : PMF X) (W : X → PMF Y)
    (d : Distortion (X := X) (Y := Y)) : ENNReal :=
  ∑ x, P x * (∑ y, (W x) y * (d x y : ENNReal))
\end{lstlisting}
The distortion values are embedded into \texttt{ENNReal}
when computing the expectation.

\subsection{Induced joint distribution}

From $P$ and a test channel $W$, we build the induced joint distribution
\[
P_{XY}(x,y)=P(x)\,W(y| x).
\]
The code again uses \texttt{bind} and \texttt{map}: “sample $x\sim P$; then sample $y\sim W(x)$; return $(x,y)$”.

\begin{lstlisting}[language=lean, frame=single]
-- Induced joint distribution
def induced_joint (P : PMF X) (W : X → PMF Y) : PMF (X × Y) :=
  P.bind (fun x => (W x).map (fun y => (x, y)))
\end{lstlisting}

\subsection{Information rate--distortion function}


In Lean, we implement the infimum using \texttt{sInf}, which here takes the infimum of a set of extended nonnegative real numbers. The set we take the infimum over is the collection of all extended nonnegative real numbers \texttt{r} such that there exists a test channel \texttt{W} with distortion at most \texttt{D} and mutual information exactly \texttt{r}.

\begin{lstlisting}[language=lean, frame=single]
-- Information rate-distortion function
noncomputable def RateDistortionFunction
    (P : PMF X) (d : Distortion (X := X) (Y := Y))
    (D : ENNReal) : ENNReal :=
  sInf { r : ENNReal | ∃ (W : X → PMF Y),
    expected_distortion_single P W d ≤ D ∧
    mutual_information (induced_joint P W) = r }
\end{lstlisting}
If no test channel satisfies the distortion constraint,
the set is empty and its infimum in \texttt{ENNReal} is $+\infty$.

This completes the formal definition of the rate--distortion function in the “test channel” form. The key point is that the Lean code mirrors the textbook definition closely: (i) test channels are functions \texttt{X → PMF Y}, (ii) the induced joint distribution is built by composition of sampling steps (\texttt{bind}/\texttt{map}), (iii) the distortion constraint is expressed as an inequality in \texttt{ENNReal}, and (iv) the rate--distortion function is an infimum of mutual informations over all channels satisfying the constraint.

\section{Formalization of the Rate Distortion Problem}
\label{sec:lean-formulation}

This section explains how we encode the \emph{formulation} of the classical rate--distortion problem. The goal is not to assume prior familiarity with Lean, but to show that the definitions closely mirror the standard mathematical ones: finite alphabets, block codes, (average) distortion, expected distortion under an i.i.d.\ source, sequences of codes, and the usual asymptotic achievability notion.

\subsection{Distortion measures}


We also parameterize the development by a block length $n$ and a codebook size $M$, together with positivity assumptions (used later to avoid degenerate divisions and empty codebooks).

\begin{lstlisting}[language=lean, frame=single]
-- blocklength and codebook size are positive
variable (n M : ℕ) (hM : 0 < M) (hn : 0 < n)
\end{lstlisting}

A (single-letter) distortion measure is a function
\[
d : \texttt{X}\times\texttt{Y}\to[0,\infty).
\]

We represent this as a function \texttt{X → Y → NNReal}, where \texttt{NNReal} denotes the nonnegative real numbers. Since the alphabets are finite, the distortion measure is bounded.

In the snippet below we introduce the type of codes at block length $n$ and size $M$:

\begin{lstlisting}[language=lean, frame=single]
-- The type of codes with blocklength n and codebook size M
def MyCode :=
  RateDistortionCode X Y n M
\end{lstlisting}

Blocks $x^n \in \texttt{X}^n$ and $y^n \in \texttt{Y}^n$ are represented as functions on \texttt{Fin n}, and codeword indices are represented by \texttt{Fin M}, which is a type representing integers in $\{0,\dots,M-1\}$.

\begin{lstlisting}[language=lean, frame=single]
-- Distortion function
def Distortion := X → Y → NNReal

-- Blocks of length n
abbrev Xblock (n : ℕ) (X : Type*) := Fin n → X
abbrev Yblock (n : ℕ) (Y : Type*) := Fin n → Y
abbrev Index (M : ℕ) := Fin M
\end{lstlisting}

\subsection{Rate--distortion codes}

A rate--distortion code consists of an encoder $f_n : \texttt{X}^n \to \{0,1,\dots,M-1\}$ and a decoder $g_n : \{0,1,\dots,M-1\} \to \texttt{Y}^n$. We use \texttt{structure} in Lean to pack the encoder and decoder together.

\begin{lstlisting}[language=lean, frame=single]
-- A rate-distortion code consists of an encoder and a decoder.
structure RateDistortionCode
  (X Y : Type*) [Fintype X] [Fintype Y]
  (n M : ℕ) where
  encoder : Xblock n X → Index M
  decoder : Index M → Yblock n Y
\end{lstlisting}

\subsection{Block distortion and code rate}

Given a single-letter distortion function $d$, we define the block distortion as the average per-letter distortion:
\[
d_n(x^n,y^n) \;=\; \frac{1}{n}\sum_{i=1}^n d(x_i,y_i).
\]
We implement this by summing the distortion over the coordinates indexed by \texttt{Fin n} and dividing by~$n$.

\begin{lstlisting}[language=lean, frame=single]
-- Block distortion
def blockDistortion
    (hn : 0 < n)
    (d : Distortion (X := X) (Y := Y))
    (x : Xblock n X) (y : Yblock n Y) : ENNReal :=
  let sum_dist := ∑ i : Fin n, (d (x i) (y i) : ENNReal)
  sum_dist / n
\end{lstlisting}

The coding rate  $R$ is defined as $\frac{1}{n}\log M$. 

\begin{lstlisting}[language=lean, frame=single]
-- Code rate of a rate-distortion code
def codeRate (hn : 0 < n) : ℝ :=
  (Real.log M) / n
\end{lstlisting}

\subsection{Discrete memoryless source model and expected distortion}

The single-letter source distribution on $X$ is represented by \texttt{PMF X}. For a block $x^n$, the i.i.d.\ probability is the product $\prod_{i=1}^n P(x_i)$; this is expressed as a finite product over \texttt{Fin n}.

\begin{lstlisting}[language=lean, frame=single]
-- Probability of a source block under the i.i.d. model
noncomputable def probBlock
    (P : PMF X) (x : Xblock n X) : ENNReal :=
  ∏ i : Fin n, P (x i)
\end{lstlisting}

We package these block probabilities into a probability mass
function \texttt{iidBlockPMF}, together with a proof that they
sum to $1$.

\begin{lstlisting}[language=lean, frame=single]
noncomputable def iidBlockPMF
    (P : PMF X) : PMF (Xblock n X) :=
  PMF.ofFintype (probBlock n P) (by
    classical
    simp only [probBlock]
    rw [← Fintype.piFinset_univ]
    rw [Finset.sum_prod_piFinset]
    simp [sum_pmf])
\end{lstlisting}

Given a code $(f_n,g_n)$, the reproduction block is $y^n = g_n(f_n(x^n))$. The expected distortion is then
\begin{align*}
&\mathbb{E}\big[d_n(X^n, g_n(f_n(X^n)))\big] \\
&= \sum_{x^n \in \texttt{X}^n} P^{n}(x^n)\, d_n\!\big(x^n, g_n(f_n(x^n))\big),
\end{align*}
where $P^{n}$ denotes the joint probability on the $n$-fold cartesian product of \texttt{X}. 
We write this as a (possibly infinite) sum \texttt{$\sum$'}. Since \texttt{X} and $\texttt{X}^n$ are finite, this coincides with an ordinary finite sum; using \texttt{$\sum$'} is convenient because it matches the measure-theoretic style used elsewhere in Mathlib.

\begin{lstlisting}[language=lean, frame=single]
-- Expected distortion of a code of length n
def expectedDistortion
    (hn : 0 < n)
    (code : RateDistortionCode X Y n M)
    (P : PMF X)
    (d : Distortion (X := X) (Y := Y)) : ENNReal :=
  ∑' x : Xblock n X, (iidBlockPMF n P x) *
    (blockDistortion n hn d x
      (code.decoder (code.encoder x)))
\end{lstlisting}

\subsection{Sequences of codes and asymptotic distortion}

A sequence of codes is indexed by $n\in\mathbb{N}$. At index $n$, we have an $(n+1,M_n)$ rate--distortion code satisfying the requirements specified in \texttt{RateDistortionCode}. Here we blocklength $n+1$, which ensures that every blocklength is positive.

\begin{lstlisting}[language=lean, frame=single]
-- A sequence of codes with positive blocklengths
structure CodeSequence
    (X Y : Type*) [Fintype X] [Fintype Y] where
  M : ℕ → ℕ
  hM : ∀ n, 0 < M n
  code : ∀ n, RateDistortionCode X Y (n + 1) (M n)
\end{lstlisting}

For a code sequence, we define the asymptotic expected distortion using a limit superior:
\[
\limsup_{n\to\infty}
\mathbb{E}\big[d_{n+1}(X^{n+1},Y^{n+1})\big].
\]
Since Mathlib uses filters to describe limits, we follow its method and use \texttt{Filter.limsup} along \texttt{atTop} to express the limit superior as $n\to\infty$.

\begin{lstlisting}[language=lean, frame=single]
-- Asymptotic expected distortion
def asymptoticDistortion
    (seq : CodeSequence X Y) (P : PMF X)
    (d : Distortion (X := X) (Y := Y)) : ENNReal :=
   Filter.limsup
    (fun n ↦
     expectedDistortion (n + 1) (seq.M n)
     (Nat.succ_pos n) (seq.code n) P d) atTop
\end{lstlisting}


\subsection{Achievability of rate--distortion pairs}

A pair $(R,D)$ is declared achievable under a pointwise
rate constraint if there exists a code sequence whose rates are all at most $R$ and whose asymptotic expected distortion is at most~$D$.
The pointwise rate condition is
\[
  \frac{\log M_n}{n+1}\leq R
  \qquad\text{for every }n\in\mathbb{N}.
\]
Here we have two constraints:
\texttt{achievesDistortion} expresses that \texttt{asymptoticDistortion} does not exceed $D$ and \texttt{hasPointwiseRateBound} expresses that the rate
of every code is bounded by $R$.

\begin{lstlisting}[language=lean, frame=single]
-- The asymptotic distortion constraint.
def achievesDistortion
    (seq : CodeSequence X Y) (P : PMF X)
    (d : Distortion (X := X) (Y := Y))
    (D : ENNReal) : Prop :=
  asymptoticDistortion seq P d ≤ D

-- Every code satisfies the prescribed rate bound.
def hasPointwiseRateBound
    (seq : CodeSequence X Y) (R : ℝ) : Prop :=
  ∀ n, codeRate (n + 1) (seq.M n)
    (Nat.succ_pos n) ≤ R
\end{lstlisting}

Then, we combine the two
constraints in \texttt{isAchievablePointwise}, which means $(R,D)$ is achievable under the pointwise rate constraint
if and only if there exists a code sequence satisfying
both constraints.

\begin{lstlisting}[language=lean, frame=single]
-- Achievability with a pointwise rate constraint.
def isAchievablePointwise
    (P : PMF X) (d : Distortion (X := X) (Y := Y))
    (R : ℝ) (D : ENNReal) : Prop :=
  ∃ seq : CodeSequence X Y,
    hasPointwiseRateBound seq R ∧
    achievesDistortion seq P d D
\end{lstlisting}

\subsection{Statements of the direct and converse theorems}

Finally, we state the rate--distortion theorem.
Here achievability refers to the pointwise rate constraint
defined above.

\begin{theorem} 
\label{thm:rate_distortion_theorem_a}
Suppose the alphabets $\texttt{X}$ and $\texttt{Y}$ are finite
and the distortion function $d$ is nonnegative and finite-valued.
For any source distribution $P$, rate $R\geq0$, and distortion
threshold $D\in[0,\infty]$, we have:
\begin{enumerate}
  \item If $R>R_I(D)$, the rate--distortion pair $(R,D)$
  is achievable.
  \item If $(R,D)$ is achievable, then $R\geq R_I(D)$.
  Consequently, if $R<R_I(D)$, the pair $(R,D)$
  is not achievable.
\end{enumerate}
\end{theorem}

We can now state the direct and converse parts of the rate--distortion theorem as follows.

\begin{lstlisting}[
  language=lean,
  frame=single,
  basicstyle=\ttfamily\scriptsize,
  breaklines=true,
  breakatwhitespace=false
]
-- Direct theorem
theorem rate_distortion_achievability
    (P : PMF X) (d : Distortion (X := X) (Y := Y))
    (D : ENNReal) (R : ℝ)
    (hRD : RateDistortionFunction P d D <
      ENNReal.ofReal R) :
    isAchievablePointwise P d R D := by
  exact isAchievablePointwise_of_rateDistortionFunction_lt
    P d D R hRD
\end{lstlisting}

\begin{lstlisting}[
  language=lean,
  frame=single,
  basicstyle=\ttfamily\scriptsize,
  breaklines=true,
  breakatwhitespace=false
]
-- Converse theorem
theorem rate_distortion_converse
    (P : PMF X) (d : Distortion (X := X) (Y := Y))
    (D : ENNReal) (R : ℝ)
    (hA : isAchievablePointwise P d R D) :
    RateDistortionFunction P d D ≤
      ENNReal.ofReal R := by
  exact rate_distortion_converse_pointwise_of_rightContinuousAt
    P d D R
    (rateDistortionRightContinuousAt_all P d D) hA
\end{lstlisting}

The distortion function is bounded because it is finite-valued and the alphabets are finite, so a separate hypothesis is no longer required.
The first part of the theorem says that if $R$ is strictly larger than the information rate distortion function, then the pair $(R,D)$ is achievable.
The second part is the converse: if $(R,D)$ is achievable, then $R\geq R_I(D)$. Consequently, if $R$ is strictly less than the information rate distortion function, then $(R,D)$ is not achievable.

The rate distortion theorem is now proved in Lean without \texttt{sorry} placeholders, and all the terms are well defined. Proofs of the rate distortion theorem can be found in standard textbooks in information theory, e.g.\ 
\cite{CoverThomas, McEliece2022, Yeung}.
The main steps of the formal proof are presented in Section~\ref{sec:lean-proof}.

\section{Proof of the Rate--Distortion Theorem}

\label{sec:lean-proof}
In this section, we explain the main steps of the formal proof. We begin with the converse, as its proof involves fewer steps. We then prove achievability for $R>R_I(D)$ with every code satisfying the prescribed rate bound. As an additional result, we prove achievability at $R=R_I(D)$ when the rate bound is required only in the limit superior. Finally, we discuss the verification and reusable components of the development.

\subsection{Converse}

In this part, we start with the finite-block inequality
\[
  R_I(D_n)\leq \frac{\log M}{n}.
\]

This follows from the feasibility of $\overline{Q}$,
convexity of mutual information under mixtures with a
fixed source marginal, independence of the source
coordinates, and the entropy bound imposed by the
codebook size. Specifically,
\[
\begin{aligned}
  R_I(D_n)
  &\leq I(\overline{Q})\\
  &\leq \frac{1}{n}\sum_{i=1}^{n}I(X_i;Y_i)\\
  &\leq \frac{1}{n}I(X^n;Y^n)
  \leq \frac{\log M}{n},
\end{aligned}
\]
where $\overline{Q}=\frac1n\sum_{i=1}^{n}Q_i$ and
$Q_i$ is the joint distribution of the source symbol
$X_i$ and its reconstruction $Y_i$.

Let $r_n$ be the rate of the code at index $n$, and
let $\Delta_n$ be its expected average distortion.
Under the pointwise achievability condition, we have
$r_n\leq R$ for every $n$ and
$\limsup_{n\to\infty}\Delta_n\leq D$.
For finite $D$, the latter condition implies that,
for any $\varepsilon>0$, there exists $N$ such that
$\Delta_n\leq D+\varepsilon$ for every $n\geq N$.
Thus, fixing any $\varepsilon>0$, we obtain the following
inequality chain for all sufficiently large $n$:
\[
\underbrace{R_I(D+\varepsilon)\leq R_I(\Delta_n)}
_{\text{Monotonicity of }R_I}
\quad
\underbrace{\leq r_n}
_{\text{Finite-block converse}}
\quad
\underbrace{\leq R}
_{\text{Pointwise rate constraint}}
\]

Using right continuity, we obtain
\[
R_I(D)
=\lim_{\varepsilon\downarrow0}R_I(D+\varepsilon)
\leq R.
\]
For $D=+\infty$, we have $R_I(D)=0$, so the conclusion
follows from $R\geq0$.

Next, we explain in detail how the Lean code corresponds
to the proof above.

For brevity, we replace the full proof code
with \texttt{...}; the full proofs are available
in the source code~\cite{RateDistortionLean}.

Firstly, the finite-block inequality is stated in Lean as follows.

\begin{lstlisting}[
  language=lean,
  frame=single,
  basicstyle=\ttfamily\scriptsize,
  breaklines=true,
  breakatwhitespace=false
]
theorem RateDistortionFunction_expectedDistortion_le_codeRate
    (hn : 0 < n)
    (code : RateDistortionCode X Y n M)
    (P : PMF X)
    (d : Distortion (X := X) (Y := Y)) :
    RateDistortionFunction P d
      (expectedDistortion n M hn code P d) ≤
      ENNReal.ofReal (codeRate n M hn) := by
  ...
\end{lstlisting}

Here \texttt{expectedDistortion} represents $D_n$,
and \texttt{codeRate} represents $\log M/n$. Thus, the theorem states $R_I(D_n)\leq \log M/n$.
The proof uses the feasibility of the average joint
distribution and the mutual-information bound
established above.

We then apply this result to a code sequence
\texttt{seq}. At index $n$, the expected average distortion $\Delta_n$
and rate $r_n$ are represented by the following
expressions, respectively.

\begin{lstlisting}[
  language=lean,
  frame=single,
  basicstyle=\ttfamily\scriptsize,
  breaklines=true,
  breakatwhitespace=false
]
-- Expected average distortion at index n
expectedDistortion (n + 1) (seq.M n)
  (Nat.succ_pos n) (seq.code n) P d

-- Rate at index n
codeRate (n + 1) (seq.M n) (Nat.succ_pos n)
\end{lstlisting}

These are the expressions used in
$$
\texttt{asymptoticDistortion}$$
and
$$\texttt{hasPointwiseRateBound},
$$
defined in Section~\ref{sec:lean-formulation}.
The implementation also uses
\texttt{asymptoticRate}, which takes the limit superior
of the code rates after conversion to \texttt{ENNReal}.

We now use the distortion constraint in the
achievability condition. For any positive distortion
slack $\varepsilon$, this constraint and the finite-block
inequality give
\[
  R_I(D+\varepsilon)\leq\limsup_{n\to\infty}r_n.
\]
This result is formalized as follows.

\begin{lstlisting}[language=lean, frame=single] 
theorem rate_distortion_sequence_converse_add
    (P : PMF X)
    (d : Distortion (X := X) (Y := Y))
    (D ε : ENNReal)
    (seq : CodeSequence X Y)
    (hD : achievesDistortion seq P d D)
    (hε : 0 < ε) :
    RateDistortionFunction P d (D + ε) ≤
      asymptoticRate seq := by
  ...
\end{lstlisting}

Here \texttt{asymptoticRate} represents the limit superior
of the code rates and takes values in \texttt{ENNReal}.

The proof uses the distortion limsup bound to obtain
an eventual bound by $D+\varepsilon$.
Monotonicity and the finite-block converse then give
an eventual lower bound on the code rates.
Mathlib's
\texttt{le\_limsup\_of\_frequently\_le}
passes this bound to their limit superior.

A pointwise rate bound implies the corresponding
limsup rate bound, which means 
\[
  \bigl(\forall n\in\mathbb{N},\ r_n\leq R\bigr)
  \quad\Longrightarrow\quad
  \limsup_{n\to\infty}r_n\leq R.
\]
This implication is formalized as follows.
\begin{lstlisting}[
  language=lean,
  frame=single,
  basicstyle=\ttfamily\scriptsize,
  breaklines=true,
  breakatwhitespace=false
]
theorem asymptoticRate_le_of_hasPointwiseRateBound
    (seq : CodeSequence X Y) (R : ℝ)
    (hR : hasPointwiseRateBound seq R) :
    asymptoticRate seq ≤ ENNReal.ofReal R := by
  ...
\end{lstlisting}
Thus, pointwise achievability implies achievability
under the asymptotic rate constraint.

Combining the result for $D+\varepsilon$ with the
asymptotic rate bound gives $R_I(D+\varepsilon)\leq R$,
formalized as follows.

\begin{lstlisting}[
  language=lean,
  frame=single,
  basicstyle=\ttfamily\scriptsize,
  breaklines=true,
  breakatwhitespace=false
]
theorem rate_distortion_converse_add
    (P : PMF X)
    (d : Distortion (X := X) (Y := Y))
    (D ε : ENNReal) (R : ℝ)
    (hε : 0 < ε)
    (hA : isAchievable P d R D) :
    RateDistortionFunction P d (D + ε) ≤
      ENNReal.ofReal R := by
  ...
\end{lstlisting}

For finite alphabets, the test-channel space is compact. Both expected distortion and mutual information are
continuous. These properties imply lower semicontinuity
of $R_I$. Together with monotonicity, this yields right
continuity at every distortion threshold.

\begin{lstlisting}[
  language=lean,
  frame=single,
  basicstyle=\ttfamily\scriptsize,
  breaklines=true,
  breakatwhitespace=false
]
def rateDistortionRightContinuousAt
    (P : PMF X)
    (d : Distortion (X := X) (Y := Y))
    (D : ENNReal) : Prop :=
  Tendsto
    (fun ε : ENNReal ↦
      RateDistortionFunction P d (D + ε))
    (nhdsWithin (0 : ENNReal) (Set.Ioi 0))
    (nhds (RateDistortionFunction P d D))

theorem rateDistortionRightContinuousAt_all
    (P : PMF X)
    (d : Distortion (X := X) (Y := Y))
    (D : ENNReal) :
    rateDistortionRightContinuousAt P d D := by
  ...
\end{lstlisting}

Filter \texttt{nhdsWithin 0 (Set.Ioi 0)} here
represents approaching zero through positive values using Mathlib's method to express limits as mentioned in Section~\ref{sec:lean-formulation}.

Finally, we use right continuity to remove the positive slack, giving $R_I(D)\leq R$.
\begin{lstlisting}[
  language=lean,
  frame=single,
  basicstyle=\ttfamily\scriptsize,
  breaklines=true,
  breakatwhitespace=false
]
theorem rate_distortion_converse_of_rightContinuousAt
    (P : PMF X)
    (d : Distortion (X := X) (Y := Y))
    (D : ENNReal) (R : ℝ)
    (hcont : rateDistortionRightContinuousAt P d D)
    (hA : isAchievable P d R D) :
    RateDistortionFunction P d D ≤
      ENNReal.ofReal R := by
  ...
\end{lstlisting}

For the pointwise converse, the implementation uses
\texttt{isAchievablePointwise\_\allowbreak
imp\_\allowbreak isAchievable}
to convert pointwise achievability to asymptotic
achievability and then applies the theorem above.
Since right continuity is proved within the development,
the final theorem requires no separate continuity
hypothesis.

\subsection{Random-coding achievability}

We now prove achievability for $R>R_I(D)$.
For finite $D$, the proof proceeds in the following steps.
\begin{enumerate}
  \item Choose a test channel satisfying the distortion
  constraint and having mutual information strictly
  below $R$.

  \item Construct a random codebook and an encoder that
  searches for a codeword satisfying the distortion
  and information-density conditions.

  \item Use probability estimates to bound the average miss probability of the random codebook.

  \item Obtain deterministic codes by averaging over
  the random-codebook distribution.

  \item Use a diagonal selection to remove the distortion
  slacks and construct a code sequence satisfying
  \[
    r_n\leq R\quad\text{for every }n,
    \qquad
    \limsup_{n\to\infty}\Delta_n\leq D.
  \]
\end{enumerate}

We next explain these steps and their formalization.

First, we choose a test channel $W$ satisfying
\[
  \mathbb{E}_{P,W}[d(X,Y)]\leq D,
  \qquad I(P,W)<R,
\]
where $I(P,W)$ is the mutual information of the joint
distribution induced by $P$ and $W$.
Such a channel exists because $R_I(D)$ is the infimum
over channels satisfying the distortion constraint
and $R_I(D)<R$.
This step is formalized as follows.

\begin{lstlisting}[
  language=lean,
  frame=single,
  basicstyle=\ttfamily\scriptsize,
  breaklines=true,
  breakatwhitespace=false
]
theorem exists_testChannel_of_RateDistortionFunction_lt_ofReal
    (P : PMF X)
    (d : Distortion (X := X) (Y := Y))
    (D : ENNReal) (R : ℝ)
    (hR : RateDistortionFunction P d D <
      ENNReal.ofReal R) :
    ∃ W : X → PMF Y,
      expected_distortion_single P W d ≤ D ∧
      mutualInformationEntropy (induced_joint P W) < R := by
  ...
\end{lstlisting}

Here \texttt{mutualInformationEntropy} is the real-valued
expression for mutual information introduced earlier.

Secondly, we construct a random codebook and its encoder.
Let $Q_Y$ be the output marginal induced by $P$ and $W$.
For blocklength $n$, we independently sample $M$
codewords, each with distribution $Q_Y^n$.
A codebook is represented as a function
\texttt{Index M → Yblock n Y}.
The corresponding probability distribution is defined
as follows.

\begin{lstlisting}[
  language=lean,
  frame=single,
  basicstyle=\ttfamily\scriptsize,
  breaklines=true,
  breakatwhitespace=false
]
-- Independently sampled block codewords
noncomputable def randomCodebookPMF
    (M n : ℕ) (Q : PMF (Yblock n Y)) :
    PMF (Index M → Yblock n Y) :=
  iidBlockPMF M Q

-- Codewords drawn from the output marginal
noncomputable def memorylessCodebookPMF
    (M n : ℕ) (P : PMF X) (W : X → PMF Y) :
    PMF (Index M → Yblock n Y) :=
  randomCodebookPMF M n
    (iidBlockPMF n (P.bind W))
\end{lstlisting}

Here \texttt{P.bind W} represents output marginal
$$\sum_x P(x)W(y\mid x)$$ 
which is $Q_Y(y)$.
The inner \texttt{iidBlockPMF n (P.bind W)} defines
the distribution of one block codeword, while
\texttt{randomCodebookPMF} defines the distribution
of $M$ independent codewords.

We then specify the condition used by the encoder.
Choose a distortion slack $\eta>0$ and an information
threshold $a$ satisfying $I(P,W)<a<R$.
A pair $(x^n,y^n)$ is good when
\[
  d_n(x^n,y^n)\leq D+\eta,
  \qquad
  \imath_n(x^n;y^n)\leq na,
\]
where $\imath_n$ is the block information density.
On the support of the joint test-channel distribution,
\[
  \imath_n(x^n;y^n)
  =\log\frac{W^n(y^n\mid x^n)}{Q_Y^n(y^n)}.
\]
These two conditions are expressed in
\texttt{goodBlockPair}.

\begin{lstlisting}[
  language=lean,
  frame=single,
  basicstyle=\ttfamily\scriptsize,
  breaklines=true,
  breakatwhitespace=false
]
noncomputable def goodBlockPair
    (hn : 0 < n) (P : PMF X) (W : X → PMF Y)
    (d : Distortion (X := X) (Y := Y))
    (distortionThreshold : ENNReal)
    (informationThreshold : ℝ)
    (x : Xblock n X) (y : Yblock n Y) : Prop :=
  blockDistortion n hn d x y ≤ distortionThreshold ∧
    memorylessInformationDensity n P W (x, y) ≤
      informationThreshold
\end{lstlisting}

In this proof, \texttt{distortionThreshold} is
$D+\eta$, and \texttt{informationThreshold} is $na$.

For a fixed codebook, the encoder selects the first
index whose codeword forms a good pair with the
source block. If no such index exists, it returns
index $0$. The decoder returns the selected codeword.
We construct this encoder using a general predicate
\texttt{good}, so that the same construction can be
reused with other conditions.

\begin{lstlisting}[
  language=lean,
  frame=single,
  basicstyle=\ttfamily\scriptsize,
  breaklines=true,
  breakatwhitespace=false
]
noncomputable def codebookEncoder
    (hM : 0 < M)
    (codebook : Index M → Yblock n Y)
    (good : Xblock n X → Yblock n Y → Prop) :
    Xblock n X → Index M := by
  classical
  intro x
  exact if h : ∃ j, good x (codebook j) then
    Fin.find (fun j ↦ good x (codebook j)) h
  else
    ⟨0, hM⟩

noncomputable def codebookCode
    (hM : 0 < M)
    (codebook : Index M → Yblock n Y)
    (good : Xblock n X → Yblock n Y → Prop) :
    RateDistortionCode X Y n M where
  encoder := codebookEncoder n M hM codebook good
  decoder := codebook
\end{lstlisting}

The assumption \texttt{hM} ensures that the fallback
index $0$ is valid.

Thirdly, we bound the average miss probability of the
random codebook. The estimate has three parts:
the good-pair probability estimate, the covering bound,
and the average miss probability bound.

For the first part, we estimate the good-pair probability
under the joint test-channel distribution.
Chebyshev bounds control deviations of the average
distortion and information density.
For the thresholds $D+\eta$ and $na$, where
$\eta>0$ and $I(P,W)<a<R$, these estimates give
\[
  \Pr_{(X^n,Y^n)\sim P^nW^n}
    \bigl((X^n,Y^n)\text{ is good}\bigr)
  \geq 1-\delta
\]
for every fixed $\delta>0$ and all sufficiently large $n$.
The corresponding theorem allows general distortion
and information thresholds.

\begin{lstlisting}[
  language=lean,
  frame=single,
  basicstyle=\ttfamily\scriptsize,
  breaklines=true,
  breakatwhitespace=false
]
theorem exists_nat_one_sub_le_goodBlockPairProbability
    (P : PMF X) (W : X → PMF Y)
    (d : Distortion (X := X) (Y := Y))
    (distortionThreshold : ENNReal)
    (informationThreshold : ℕ → ℝ)
    (εd εi δ : ℝ)
    (hεd : 0 < εd) (hεi : 0 < εi)
    (hδ : 0 < δ)
    (hdistortionThreshold : distortionThreshold ≠ ⊤)
    (hdistortionSlack :
      (expected_distortion_single P W d).toReal + εd ≤
        distortionThreshold.toReal)
    (hinformationSlack : ∀ n : ℕ, (n : ℝ) *
      (mutualInformationEntropy (induced_joint P W) + εi) ≤
        informationThreshold n) :
    ∃ N : ℕ, 0 < N ∧ ∀ n : ℕ, N ≤ n → ∀ hn : 0 < n,
      1 - ENNReal.ofReal δ ≤
        goodBlockPairProbability n hn P W d
          distortionThreshold (informationThreshold n) := by
  ...
\end{lstlisting}

Here \texttt{goodBlockPairProbability} is the joint
good-pair probability.

Let $q_n(x^n)$ be the conditional good-pair probability
under $W^n(\cdot\mid x^n)$.
Since the joint good-pair probability is the average
of $q_n(X^n)$, the preceding estimate implies
\[
  \Pr_{X^n\sim P^n}\bigl(q_n(X^n)<1/2\bigr)
  \leq 2\delta.
\]
This identifies the source blocks on which we can
apply the covering estimate.

For the second part, we derive a covering bound.
Let $p_n(x^n)$ be the probability that a codeword
drawn from $Q_Y^n$ forms a good pair with $x^n$.
For source blocks with positive probability, a
change-of-measure argument using the information-density
condition gives
\[
  p_n(x^n)\geq e^{-na}q_n(x^n).
\]
When $q_n(x^n)\geq1/2$, independence of the $M$ codewords
therefore gives
\[
\begin{aligned}
&\phantom{=}  \Pr(\text{all codewords fail}\mid x^n) \\
  &=(1-p_n(x^n))^M\\
  &\leq e^{-Mp_n(x^n)}
  \leq \exp\left(-\frac{M}{2}e^{-na}\right).
\end{aligned}
\]
Choose $M=\lfloor e^{nR}\rfloor$.
For sufficiently large $n$, $M\geq e^{nR}/2$, so
\[
  \Pr(\text{all codewords fail}\mid x^n)
  \leq \exp\left(-\frac14 e^{n(R-a)}\right).
\]
The following theorem combines the change-of-measure,
independence, and codebook-size estimates.

\begin{lstlisting}[
  language=lean,
  frame=single,
  basicstyle=\ttfamily\tiny,
  identifierstyle=\ttfamily\tiny,
  columns=fullflexible,
  breaklines=true,
  breakatwhitespace=false,
  breakautoindent=false,
  breakindent=1em
]
theorem codebookMissProbability_le_exp_neg_quarter_exp_rateGap_of_conditionalGood
    (n : ℕ) (rate threshold : ℝ)
    (hlarge : 2 ≤ Real.exp ((n : ℝ) * rate))
    (hn : 0 < n) (P : PMF X) (W : X → PMF Y)
    (d : Distortion (X := X) (Y := Y))
    (distortionThreshold : ENNReal)
    (x : Xblock n X)
    (hx : iidBlockPMF n P x ≠ 0)
    (hhalf : 1 / 2 ≤
      conditionalGoodBlockPairProbability n hn P W d
        distortionThreshold ((n : ℝ) * threshold) x)
    [DecidablePred
      (goodBlockPair n hn P W d distortionThreshold
        ((n : ℝ) * threshold) x)] :
    codebookMissProbability (exponentialCodebookSize n rate)
      (iidBlockPMF n (P.bind W))
      (goodBlockPair n hn P W d distortionThreshold
        ((n : ℝ) * threshold) x) ≤
      ENNReal.ofReal (Real.exp (-(1 / 4 *
        Real.exp ((n : ℝ) * (rate - threshold))))) := by
  ...
\end{lstlisting}

Here \texttt{rate} and \texttt{threshold} represent
$R$ and $a$.
The assumptions \texttt{hx} and \texttt{hhalf} require
a positive-probability source block with conditional
good-pair probability at least $1/2$.
The assumption \texttt{hlarge} provides the lower bound
on the rounded codebook size.

For the third part, we average over the source blocks.
Those with $q_n(x^n)<1/2$ have probability at most
$2\delta$, and their miss probability is at most $1$.
On the remaining positive-probability source blocks,
the preceding exponential bound applies.
Thus,
\[
  \mathbb{E}_{\mathcal{C}}
    [p_{\mathrm{miss}}(\mathcal{C})]
  \leq
  2\delta+\exp\left(-\frac14 e^{n(R-a)}\right),
\]
where $p_{\mathrm{miss}}(\mathcal{C})$ is the source
probability of finding no good codeword in codebook
$\mathcal{C}$.
The final average bound is formalized as follows.

\begin{lstlisting}[
  language=lean,
  frame=single,
  basicstyle=\ttfamily\tiny,
  identifierstyle=\ttfamily\tiny,
  columns=fullflexible,
  breaklines=true,
  breakatwhitespace=false,
  breakautoindent=false,
  breakindent=1em
]
theorem averageCodebookMissProbability_le_two_mul_add_exp_neg_quarter_exp_rateGap
    (n : ℕ) (rate threshold : ℝ)
    (hlarge : 2 ≤ Real.exp ((n : ℝ) * rate))
    (hn : 0 < n) (P : PMF X) (W : X → PMF Y)
    (d : Distortion (X := X) (Y := Y))
    (distortionThreshold δ : ENNReal)
    [∀ x : Xblock n X, DecidablePred
      (goodBlockPair n hn P W d distortionThreshold
        ((n : ℝ) * threshold) x)]
    [DecidablePred (fun x : Xblock n X => 1 / 2 ≤
      conditionalGoodBlockPairProbability n hn P W d
        distortionThreshold ((n : ℝ) * threshold) x)]
    (hsource : 1 - 2 * δ ≤
      codewordHitProbability (iidBlockPMF n P)
        (fun x => 1 / 2 ≤
          conditionalGoodBlockPairProbability n hn P W d
            distortionThreshold ((n : ℝ) * threshold) x)) :
    averageCodebookMissProbability
      (exponentialCodebookSize n rate)
      (iidBlockPMF n P) (iidBlockPMF n (P.bind W))
      (goodBlockPair n hn P W d distortionThreshold
        ((n : ℝ) * threshold)) ≤
      2 * δ + ENNReal.ofReal (Real.exp (-(1 / 4 *
        Real.exp ((n : ℝ) * (rate - threshold))))) := by
  ...
\end{lstlisting}

The assumption \texttt{hsource} expresses that source
blocks with conditional good-pair probability at least
$1/2$ have probability at least $1-2\delta$.
The parameter $\delta$ uses \texttt{ENNReal} to match
the probability values and is finite here, as it comes
from the real tolerance chosen earlier.
And since $a<R$, the exponential term tends to zero.
This completes the probability estimate used to
obtain deterministic codes in the next step.

Fourthly, we obtain deterministic codes
from these probability estimates.

Let $d_{\max}$ be an upper bound on the single-letter
distortion. If a good codeword is found, the average
distortion is at most $D+\eta$; otherwise, it is at
most $d_{\max}$.
Hence, for a fixed codebook $\mathcal{C}$, its expected
average distortion satisfies
\[
  \Delta_n(\mathcal{C})
  \leq D+\eta+d_{\max}p_{\mathrm{miss}}(\mathcal{C}).
\]
Averaging over the random-codebook distribution gives
a bound on the average expected distortion.
At least one deterministic codebook has expected
distortion no greater than this average.

For every $\eta>0$ and $0<\delta\leq1$, the exponential
term above is at most $\delta$ for all sufficiently
large $n$. We therefore obtain deterministic codes
satisfying
\[
  r_n\leq R,
  \qquad
  \Delta_n\leq D+\eta+3d_{\max}\delta,
\]
where $r_n$ and $\Delta_n$ are the rate and expected
average distortion of the selected length-$n$ code.
This result is formalized as follows.

\begin{lstlisting}[
  language=lean,
  frame=single,
  basicstyle=\ttfamily\scriptsize,
  breaklines=true,
  breakatwhitespace=false
]
theorem exists_nat_forall_exists_testChannelCode_bound
    (P : PMF X) (W : X → PMF Y)
    (d : Distortion (X := X) (Y := Y))
    (D : ENNReal) (R : ℝ)
    (hDtop : D ≠ ⊤)
    (hW : expected_distortion_single P W d ≤ D)
    (hI : mutualInformationEntropy (induced_joint P W) < R)
    (η δ : ℝ)
    (hη : 0 < η) (hδ : 0 < δ) (hδle : δ ≤ 1) :
    ∃ N : ℕ, ∀ n : ℕ, N ≤ n → ∃ hn : 0 < n,
      ∃ code : RateDistortionCode X Y n
          (exponentialCodebookSize n R),
        codeRate n (exponentialCodebookSize n R) hn ≤ R ∧
        expectedDistortion n (exponentialCodebookSize n R)
          hn code P d ≤
            D + ENNReal.ofReal η +
              (distortionBound d : ENNReal) *
                (3 * ENNReal.ofReal δ) := by
  ...
\end{lstlisting}

Here \texttt{exponentialCodebookSize n R} represents
$\lfloor e^{nR}\rfloor$, and \texttt{distortionBound d}
provides $d_{\max}$.
The conclusion supplies a code at every blocklength
$n\geq N$ for each fixed choice of $\eta$ and $\delta$.

Finally, we construct a code sequence and remove
the distortion slacks.
Choose $\eta_k>0$ and $0<\delta_k\leq1$ such that
$\eta_k,\delta_k\to0$.
For each precision level $k$, the finite-block result
supplies codes beyond a blocklength threshold $N_k$.
A diagonal selection lets the precision level increase
with the blocklength, so that
\[
  \eta_k+3d_{\max}\delta_k\longrightarrow0.
\]
Constant-reconstruction codes fill the finitely many
initial blocklengths.
The resulting sequence satisfies
\[
  r_n\leq R\quad\text{for every }n,
  \qquad
  \limsup_{n\to\infty}\Delta_n\leq D.
\]
The corresponding code-sequence theorem is:

\begin{lstlisting}[
  language=lean,
  frame=single,
  basicstyle=\ttfamily\scriptsize,
  breaklines=true,
  breakatwhitespace=false
]
theorem exists_codeSequence_testChannel_achieves
    (P : PMF X) (W : X → PMF Y)
    (d : Distortion (X := X) (Y := Y))
    (D : ENNReal) (R : ℝ)
    (hDtop : D ≠ ⊤)
    (hW : expected_distortion_single P W d ≤ D)
    (hI : mutualInformationEntropy (induced_joint P W) < R) :
    ∃ seq : CodeSequence X Y,
      hasPointwiseRateBound seq R ∧
      asymptoticDistortion seq P d ≤ D := by
  ...
\end{lstlisting}

The proof uses
\texttt{exists\_diagonalCodeSequence\_\allowbreak
of\_eventually\_codes}
to construct the sequence and
\texttt{asymptoticDistortion\_\allowbreak
le\_of\_diagonal\_error}
to establish its distortion bound.
The conclusion is the pointwise achievability condition
introduced in Section~\ref{sec:lean-formulation}.

When $D=+\infty$, there is no distortion restriction.
The chosen test channel supplies a nonempty reproduction
alphabet, and $R>I(P,W)\geq0$.
A constant-reconstruction code with a single codeword
therefore satisfies the rate bound at every blocklength.
This completes the direct theorem.

\subsection{Achievability at the equality rate}

We now consider $R=R_I(D)$ when $R_I(D)$ is finite.
We use the same pointwise rate constraint as in the
strict-rate theorem. The proof proceeds as follows.
\begin{enumerate}
  \item For $R>0$, mix an optimal test channel with an
  independent reproduction channel to obtain mutual
  information strictly below $R$, allowing a small
  increase in distortion.

  \item Apply the random-coding estimates at the fixed
  rate $R$, and use the diagonal selection from the
  preceding subsection to remove the distortion errors.

  \item For $R=0$, use a constant reconstruction obtained
  from an optimal test channel with zero mutual information.
\end{enumerate}

First, suppose that $R>0$ and $D$ is finite. Since the infimum defining $R_I(D)$ is attained, we can choose an optimal test channel $W^*$ satisfying
\[
  \mathbb{E}_{P,W^*}[d(X,Y)]\leq D,
  \qquad I(P,W^*)=R.
\]
Let $Q_Y=P W^*$ be its output marginal, and define
an independent reproduction channel by
$W_0(y\mid x)=Q_Y(y)$.
For $0<t<1$, consider the mixture
\[
  W_t(y\mid x)
  =(1-t)W^*(y\mid x)+tQ_Y(y).
\]
Convexity of mutual information with a fixed source
distribution gives
\[
  I(P,W_t)
  \leq (1-t)I(P,W^*)+tI(P,W_0)
  =(1-t)R<R.
\]
If $d_{\max}$ is an upper bound on the distortion,
linearity of expected distortion gives
\[
  \mathbb{E}_{P,W_t}[d(X,Y)]
  \leq D+t d_{\max}.
\]
Thus, for any $\eta>0$, choosing $t$ sufficiently small
provides a test channel with distortion at most
$D+\eta$ and mutual information strictly below $R$.
This construction is formalized in the following theorem,
which also applies when $R_I(D)\leq R$.

\begin{lstlisting}[
  language=lean,
  frame=single,
  basicstyle=\ttfamily\scriptsize,
  breaklines=true,
  breakatwhitespace=false
]
theorem exists_testChannel_strict_rate_with_distortion_slack
    (P : PMF X)
    (d : Distortion (X := X) (Y := Y))
    (D : ENNReal) (R : ℝ)
    (hR : 0 < R)
    (hRD : RateDistortionFunction P d D ≤
      ENNReal.ofReal R)
    (η : ℝ) (hη : 0 < η) :
    ∃ W : X → PMF Y,
      expected_distortion_single P W d ≤
        D + ENNReal.ofReal η ∧
      mutualInformationEntropy (induced_joint P W) < R := by
  ...
\end{lstlisting}

The proof uses \texttt{channelMix} to construct the mixture
and \texttt{P.bind W} to obtain the output marginal.
The parameter $\eta$ controls the additional distortion.

Next, we apply the random-coding estimates from the
preceding subsection at the fixed rate $R$.
Set $s_k=1/(k+1)$.
For each $k$, the mixed channel has distortion at most
$D+s_k$ and mutual information strictly below $R$.
Using $s_k$ for both error parameters in the
random-coding estimates gives codes at all sufficiently
large blocklengths satisfying
\[
  r_n\leq R,
  \qquad
  \Delta_n\leq D+2s_k+3d_{\max}s_k.
\]
The additional $s_k$ comes from the channel mixture.
Since the total distortion error tends to zero,
the same diagonal selection produces a code sequence with
\[
  r_n\leq R\quad\text{for every }n,
  \qquad
  \limsup_{n\to\infty}\Delta_n\leq D.
\]

For the zero-rate case, an optimal test channel has
zero mutual information, so its induced joint distribution
is the product of its marginals.
Averaging over the reproduction marginal then gives
a symbol $y_0$ such that
\[
  \sum_x P(x)d(x,y_0)\leq D.
\]
Using $y_0$ at every coordinate gives codes with
$M_n=1$, rate zero, and expected average distortion
at most $D$. This case is formalized as follows.

\begin{lstlisting}[
  language=lean,
  frame=single,
  basicstyle=\ttfamily\scriptsize,
  breaklines=true,
  breakatwhitespace=false
]
theorem rate_distortion_zero_rate_achievable_pointwise
    (P : PMF X)
    (d : Distortion (X := X) (Y := Y))
    (D : ENNReal)
    (hzero : RateDistortionFunction P d D = 0) :
    isAchievablePointwise P d 0 D := by
  ...
\end{lstlisting}

The case $D=+\infty$ is also handled by constant
reconstruction codes. Combining these cases gives
the equality-rate theorem.

\begin{lstlisting}[
  language=lean,
  frame=single,
  basicstyle=\ttfamily\scriptsize,
  breaklines=true,
  breakatwhitespace=false
]
theorem rate_distortion_achievability_at_boundary
    (P : PMF X)
    (d : Distortion (X := X) (Y := Y))
    (D : ENNReal)
    (hfinite : RateDistortionFunction P d D ≠ ⊤) :
    isAchievablePointwise P d
      (RateDistortionFunction P d D).toReal D := by
  exact rate_distortion_achievability_at_boundary_pointwise
    P d D hfinite
\end{lstlisting}

The assumption \texttt{hfinite} ensures that
\texttt{toReal} represents $R_I(D)$ as a finite real rate.
Both the strict-rate and equality-rate results now use
\texttt{isAchievablePointwise}.

\subsection{Verification and reusable components}

The complete Lean formalization, including all definitions,
proofs, and verification scripts, is available in the
GitHub repository~\cite{RateDistortionLean}.

To make effective use of AI assistance and reduce the
risk of mismatches with the intended mathematics,
we first wrote the definitions needed by the main
theorem and its statement, using \texttt{sorry} as a
temporary placeholder for the proof.
We carefully checked the definitions, hypotheses,
quantifiers, and conclusion against their mathematical
counterparts before using AI assistance to complete
the proof. Lean then checked the resulting proof.

When formalizing the main theorem at $R=R_I(D)$,
we found that AI used the weaker notion of achievability
\texttt{isAchievable} rather than
\texttt{isAchievablePointwise}, which is used for
$R>R_I(D)$.
After carefully checking the statement, we revised it
and completed the proof under the pointwise rate constraint.
This example show that Lean verification alone
does not ensure alignment with the intended
mathematics~\cite{Bastounis2026}, and that theorem
statements still require careful review.

The Lean module compiles without \texttt{\small sorry}
placeholders. We also inspect the axioms used by the
main results with \texttt{\small \#print axioms}.

\begin{lstlisting}[
  language=lean,
  frame=single,
  basicstyle=\ttfamily\scriptsize,
  breaklines=true,
  breakatwhitespace=false
]
#print axioms rate_distortion_achievability
#print axioms rate_distortion_converse
#print axioms rate_distortion_achievability_at_boundary
\end{lstlisting}

These checks report only \texttt{\small propext},
\texttt{\small Classical.choice}, and \texttt{\small Quot.sound},
with no dependency on \texttt{\small sorry\allowbreak Ax}.

For the purpose of building formal infrastructure for
information theory in Lean, the development reuses
Mathlib's existing definitions and results wherever possible,
including probability mass functions, finite sums and
products, filters, and topological results.

We provide reusable definitions and lemmas for
finite-alphabet information measures, optimization over
test channels, and coding arguments. These include
information inequalities and continuity results for
\texttt{\small shannon\allowbreak Entropy} and
\texttt{\small mutual\_\allowbreak information},
compactness of \texttt{\small PMF} on finite alphabets, and
the existence of optimal test channels under feasible
distortion constraints in
\texttt{\small Rate\allowbreak Distortion\allowbreak Function\_\allowbreak attained}.
We also establish lower semicontinuity and right
continuity of $R_I$.

\begingroup
\rightskip=0pt plus 1fil\relax
For coding arguments,
\texttt{\small random\allowbreak Codebook\allowbreak PMF}
constructs random codebooks with independently sampled
codewords, and
\texttt{\small codebook\allowbreak Encoder}
uses a general \texttt{\small good} predicate.
The diagonal construction produces a
\texttt{\small Code\allowbreak Sequence}
from finite-block estimates to obtain asymptotic guarantees.
These components allow later developments to reuse the
probability and limit arguments established here.
The aim is to expand the formal foundations available
for information theory and support the formalization
of further results.\par
\endgroup

\section{Concluding Remarks}
\label{sec:conclusion}
Terence Tao in his ICM2026 lecture~\cite{Tao2026ICM}
outlined a prospective workflow for future mathematics:
human researchers specify problems in a formal language,
AI systems solve them and produce machine-certified proofs,
and mathematicians subsequently interpret these proofs
and extract their conceptual content.
Formalization thus plays a pivotal role in this process.
In order to apply AI-assisted methodology in information
theory, we need to first formalize the basic results
before proceeding to open conjectures.

Our infrastructure for rate--distortion theory makes
the objects, assumptions, and conventions explicit,
including the treatment of zero-probability events,
extended nonnegative real values, and feasibility constraints.
We prove the direct and converse parts of the
finite-alphabet rate--distortion theorem in Lean and
provide reusable definitions and lemmas built on Mathlib.
Future work may extend these foundations to
rate--distortion--perception tradeoffs and their applications
in machine learning and communication.
This paper contributes to the construction of formal
foundations on which AI systems can generate
machine-checkable proofs while preserving the conceptual
clarity valued by human researchers.





\bibliographystyle{IEEEtran_unsrt} 
\bibliography{Ref}

@STRING{ISITA = {{IEEE} Int. Symp. Information Theory and Its Applications}}

@book{Berger71,
	author    = "T. Berger",
	title     = "Rate distortion theory",
	publisher = "Prentice-Hall",
	address   = "Englewood Cliff",
	year 	  = 1971
}

@book{Viterbi2009,
	author    = "A. J. Viterbi and J. K. Omura",
	title     = "Principles of Digital Communication and Coding",
	publisher = "Dover Publications",
	year 	  = 2009
}

@book{McEliece2022,
	author    = "R. J. McEliece",
	title     = "The theory of Information and Coding",
	publisher = "Cambridge University Press",
	edition   = "second",
	year 	  = 2002
}

@book{CoverThomas,
	author    = "T. M. Cover and J. A. Thomas",
	title     = "Elements of information theory",
	publisher = "John Wiley \& Sons",
	edition   = "second",
	year 	  = 2006
}

@book{Yeung,
	author    = "R. W. Yeung",
	title     = "Information theory and network coding",
	publisher = "Springer",
	year 	  = 2008
}

@inproceedings{Mhamdi2011,
	author    = "T. Mhamdi and O. Hasan and S. Tahar",
	title     = "Formalization of entropy measures in {HOL}",
	booktitle = "Int. Conf. on Interactive Theorem Proving (ITP)",
	series   = "Lecture notes in computer science",
	volume   = 6869,
	month     = nov,
	pages     = "233-248",
	year      = 2011
}

@inproceedings{Affeldt2012,
	author    = {Reynald Affeldt and Manabu Hagiwara and Jonas Senizergues},
	title     = {Formalization of {S}hannon's Theorems in {SSReflect}-{Coq}},
	booktitle = {Interactive Theorem Proving (ITP)},
	series    = {Lecture Notes in Computer Science},
	publisher = {Springer},
	address   = {Princeton, NJ},
	month     = aug,
	year      = 2012
}

@inproceedings{Nakano2016,
	author    = "K. Nakano and M. Hagiwara",
	title     = "Formalization of binary symmetric erasure channel based on infotheo",
	booktitle = ISITA,
	address   = "Monterey, CA",
	month     = nov,
	year      = 2016
}

@article{Affeldt2014,
	author    = "Reynald Affeldt and Manabu Hagiwara and Jonas Senizergues",
title     = "Formalization of {S}hannon's Theorems",
journal =   "Journal of Automated Reasoning",
volume    = 53,
pages     = "63--103",
year      = 2014
}

@misc{CoqInfotheo,
	author       = {Reynald Affeldt and others},
	title        = {Infotheo: A {C}oq Library for Information Theory},
	howpublished = {\url{https://github.com/affeldt-aist/infotheo}}
}

@mastersthesis{Luccioli_thesis,
	title        = "Formalizing information theory in {L}ean 4: $f$-divergences, hypothesis testing and the data processing inequality",
	author       = "L. Luccioli",
	year         = 2024,
	school       = "University di Bologna",
	type         = {Master thesis}
}

@article{gowers2025conjecture,
	title={On a conjecture of {M}arton},
	author={Gowers, William Timothy and Green, Ben and Manners, Freddie and Tao, Terence},
	journal={Annals of Mathematics},
	volume={201},
	number={2},
	pages={515--549},
	year= 2025
}

@misc{kasaura2026lean,
	title={Lean-{Q}uantum: Toward {AI}-Assisted Formalization of Quantum Information},
	author={Kazumi Kasaura and Kei Tsukamoto and Kento Mori and Risa Mizuno and Takahiro Namatame and Yuta Oriike and Masaya Taniguchi and Sho Sonoda and Hayata Yamasaki},
	howpublished = {\url{https://arxiv.org/abs/2607.05492}},
	month = jul,
	year=2026
}

@misc{AnthropicFLT2026,
  author       = "Anthropic",
  title        = "Formalizing {F}ermat's Last Theorem",
  year         = 2026,
  month        = sep,
  howpublished = {\href{https://www.anthropic.com/research/formalizing-fermats-last-theorem}{Link to webpage}}
}

@misc{OpenAINavierStokes2026,
  author       = {Open{AI}},
  title        = {On the {N}avier--{S}tokes Millennium Prize Problem},
  year         = 2026,
  month        = sep,
  howpublished = {\href{https://openai.com/index/navier-stokes-solution/}{Link to webpage}}
}

@misc{Huang2026,
  author       = {Mian Huang and Yanxiao Liu and Yi Liu},
  title        = {Sub-optimality of {M}arton's Inner Bound
                  for the Two-Receiver Broadcast Channel},
  month        = aug,
  year         = 2026,
  howpublished = {\href{https://arxiv.org/abs/2608.19869}{arXiv:2608.19869 [cs.IT]}}
}

@misc{Tandon2026,
  author       = {Ravi Tandon},
  title        = {Strengthening Recursive Constructions for Zero-Error {S}hannon Capacity},
  month        = aug,
  howpublished = {\href{https://arxiv.org/abs/2608.30273}{arXiv:2608.30273 [cs.IT]}},
  year         = 2026
}

@misc{Itty2026improved,
	author       = {Nathaniel Itty and Christopher D. Rosin and Chase Carstensen and Daniel Reichman},
	title        = {Improved lower bounds for the {S}hannon capacity of odd cycles},
	month        = jul,
	howpublished = {\href{https://arxiv.org/abs/2607.21517}{arXiv:2607.21517 [cs.IT]}},
	year         = 2026
}

@misc{Protti2026Shannon,
	author       = {Matthew Protti},
	title        = {An improved lower bound for the {S}hannon capacity of 
	{$C_{11}$}},
	month        = sep,
	howpublished = {\href{https://github.com/matthewprotti/c11-shannon-capacity-lower-bound}{link to github}},
	year         = 2026
}

@misc{YangDongN4Code,
  author       = {Shenghao Yang and Yanyan Dong},
  title        = {A machine-checked proof of the {D}ong-{Y}ang classification of optimal $(n,4)$ binary codes for {BSC}s},
  month        = sep,
  howpublished = {\href{https://arxiv.org/abs/2609.10579}{arXiv:2609.10579v1 [math.HO]}},
  year         = 2026
}

@misc{PFRLean,
	author = "T. Tao",
	title = "The Polynomial {F}reiman-{R}uzsa Conjecture",
	howpublished = {\url{https://teorth.github.io/pfr/}},
	year =  2025
}

@misc{zhang2026empirical,
	title={{AI4SLT}: Empirical Processes in Lean 4 for Formal Statistical Learning Theory},
	author={Yuanhe Zhang and Jason D. Lee and Fanghui Liu},
	howpublished = {\href{https://arxiv.org/abs/2602.02285v2}{arXiv:2602.02285v2 [cs.LG]}},
	note = "Accepted by ICML2026",
	month = jun,
	year= 2026
}

@misc{sonoda2026Lean,
	title={{L}ean Formalization of Generalization Error Bound by {R}ademacher Complexity and {D}udley's Entropy Integral},
	author={Sho Sonoda and Kazumi Kasaura and Yuma Mizuno and Kei Tsukamoto and Naoto Onda},
	howpublished = {\url{https://arxiv.org/abs/2503.19605}},
	note = "Accepted by ITP2026",
	month = may,
	year= 2026
}

@misc{Degenne2026Lean,
	title={{L}ean Machine Learning -- the {L}ean library for machine learning research},
	author={R\'{e}my Degenne and Paulo Rauber},
	howpublished = {\href{https://leanmachinelearning.org/}{Link to github}},
	year= 2026
}

@misc{Tao2026ICM,
	title={Mathematics in the age of {AI}},
	author={Terence Tao},
	howpublished = {\url{https://teorth.github.io/tao-web/slides/age-of-ai-icm-2026.pdf}},
	month = jul,
	year= 2026
}

@misc{OpenAI10Advances,
	title={Ten advances in mathematics and theoretical computer science},
	author= {Open{AI}},
	howpublished = {\href{https://cdn.openai.com/pdf/ten-proofs-oai.pdf}{Link to pdf}},
	month = aug,
	year= 2026
}

@misc{RateDistortionLean,
  author       = {Fubin Yan and Ruxuan Dong and Ruoxuan Shi and Duolei Wang and Kenneth W. Shum},
  title        = {Rate Distortion Theory in Lean 4},
  howpublished = {\url{https://github.com/fubinyan/rate_distortion_lean4}},
  year         = {2026}
}

@misc{Pichler2026,
	author       = {Georg Pichler},
	title        = {Three Conjectures on Binary Channels for the Doubly Symmetric Binary Source},
	month        = sep,
	howpublished = {\href{https://arxiv.org/abs/2609.27991v1}{arXiv:2609.27991 [cs.IT]}},
	year         = 2026
}

@misc{Durcik2026weak,
	author       = {Polona Durcik and Marco Fraccaroli and Joris Roos},
	title        = {A weak {H}ellinger inequality for noisy {B}oolean channels},
	month        = sep,
	howpublished = {\href{https://arxiv.org/abs/2609.28534}{	arXiv:2609.28534 [cs.IT]}},
	year         = 2026
}

@misc{Soysal2026new,
	author       = {{\"{O}}zg\"{u}r Soysal},
	title        = {A New Upper Bound on the Binary Deletion Channel Capacity},
	howpublished = {\href{https://arxiv.org/abs/2609.13351}{arXiv:2609.13351 [cs.IT]}},
	month        = sep,
	year         = 2026
}

@misc{Bastounis2026,
  author        = {Bastounis, Alexander and Circelli, Fabian
                   and Hansen, Anders C.},
  title         = {{Navier-Stokes} lost in translation:
                   Why {Lean} verification of {AI} autoformalisation
                   does not guarantee correct natural language proofs},
  year          = {2026},
  eprint        = {2610.08144},
  archivePrefix = {arXiv},
  primaryClass  = {math.AP},
  doi           = {10.48550/arXiv.2610.08144},
  url           = {https://arxiv.org/abs/2610.08144}
}

@misc{OAI:The-circulant-Hadamard-conjecture-September-23-2026,
  author = {Open{AI}},
  title = {The circulant {H}adamard conjecture},
  howpublished = {OpenAI Math Release preprint
\href{https://github.com/openai/math/blob/main/preprints/The-circulant-Hadamard-conjecture-September-23-2026/paper.pdf}{OAI:The-circulant-Hadamard-conjecture-September-23-2026}},
  year = 2026
}
\end{document}